\documentclass[%
 aip,
 jmp,%
 amsmath,amssymb,
 reprint
]{revtex4-1}

\usepackage{graphicx}
\usepackage{dcolumn}
\usepackage{bm}

\begin{document}


\title[]{Two-fluid simulations of driven reconnection in the Mega-Ampere Spherical Tokamak}

\author{A. Stanier}
 \email{stanier@jb.man.ac.uk}
\author{P. Browning}%
\author{M. Gordovskyy}
\affiliation{Jodrell Bank Centre for Astrophysics, University of Manchester, Manchester, M13 9PL, UK}

\author{K. G. McClements}
\affiliation{EURATOM/CCFE Fusion Association, Culham Science Centre, Abingdon, Oxfordshire OX14 3DB, UK}

\author{M. P. Gryaznevich}
\affiliation{EURATOM/CCFE Fusion Association, Culham Science Centre, Abingdon, Oxfordshire OX14 3DB, UK}
\affiliation{Present affiliation: Imperial College of Science and Technology, London SW7 2AZ, UK }

\author{V. S. Lukin}
\affiliation{Space Science Division, Naval Research Laboratory, Washington, DC 20375, USA }

\date{\today}

\begin{abstract}
In the merging-compression method of plasma start-up, two flux-ropes with parallel toroidal current are formed around in-vessel poloidal field coils, before merging to form a spherical tokamak plasma. This start-up method, used in the Mega-Ampere Spherical Tokamak (MAST), is studied as a high Lundquist number and low plasma-beta magnetic reconnection experiment. 

In this paper, 2D fluid simulations are presented of this merging process in order to understand the underlying physics, and better interpret the experimental data. These simulations examine the individual and combined effects of tight-aspect ratio geometry and two-fluid physics on the merging. The ideal self-driven flux-rope dynamics are coupled to the diffusion layer physics, resulting in a large range of phenomena. For resistive MHD simulations, the flux-ropes enter the sloshing regime for normalised resistivity $\eta \lesssim 10^{-5}$. In Hall-MHD three regimes are found for the qualitative behaviour of the current sheet, depending on the ratio of the current sheet width to the ion-sound radius. These are a stable collisional regime, an open X-point regime, and an intermediate regime that is highly unstable to tearing-type instabilities. 

In toroidal axisymmetric geometry, the final state after merging is a MAST-like spherical tokamak with nested flux-surfaces. It is also shown that the evolution of simulated 1D radial density profiles closely resembles the Thomson scattering electron density measurements in MAST. An intuitive explanation for the origin of the measured density structures is proposed, based upon the results of the toroidal Hall-MHD simulations. 

\end{abstract}

\maketitle
\section{\label{intro}Introduction}

The Mega-Ampere Spherical Tokamak (MAST)\cite{lloyd11} has demonstrated promising confinement scalings while operating at higher equilibrium plasma-$\beta$ values than conventional tokamaks. It has been suggested that the Spherical Tokamak (ST) magnetic confinement concept should be developed further, towards a ST Power Plant~\cite{voss02} or ST based Component Test Facility (CTF)\cite{peng05}. In these machines significant neutron shielding would be required for the toroidal field coils at the central column, leaving little space for a central solenoid. An attractive option is to remove the central solenoid and to achieve plasma formation and current drive through other methods. Towards this goal several non-solenoidal start-up methods have been investigated on a number of devices, including the use of radio-frequency waves~\cite{shiraiwa04,gryaznevich06}, co-axial and DC helicity injection~\cite{raman2010,battaglia2011}, and flux-rope merging start-up via poloidal field coil induction~\cite{sykes01,yamada2010}. 

\begin{figure}
\includegraphics[width=0.425\textwidth]{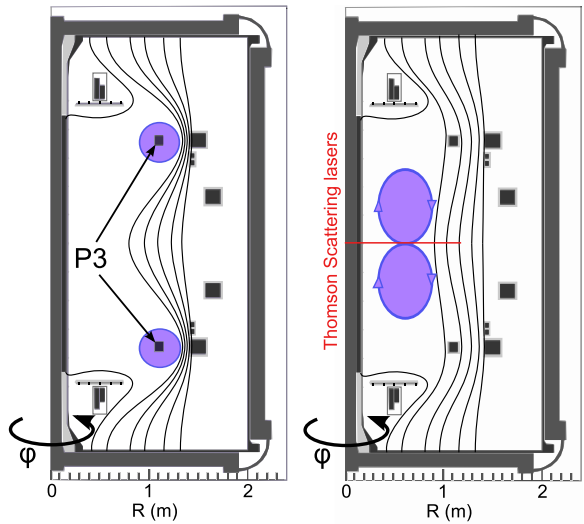}
\caption{\label{fig:cartoon1} A cartoon of the merging-compression process within the MAST vacuum vessel. Cross-sections of the poloidal field coils are black squares, where the P3 poloidal field coils are labelled. The red line indicates the position of the Nd:YAG Thomson Scattering laser diagnostic. The purple rings are the flux-ropes with arrows indicating the poloidal field direction.}
\end{figure}

The merging-compression start-up method, first performed on the Small Tight Aspect Ratio Tokamak~\cite{start92}, is now routinely used on MAST~\cite{sykes01}. After gas filling and ramp-up of currents in toroidal and poloidal field coils, to supply the vacuum field, the current in the pair of P3 poloidal field coils (see Figure~\ref{fig:cartoon1}) is ramped back down towards zero on a millisecond timescale. This causes breakdown and induces toroidal current rings, or co-helicity flux-ropes, in the plasma surrounding the P3 coils. When the parallel toroidal plasma current within the flux-ropes becomes greater than the current in the respective P3 coils, the mutual attraction between the flux-ropes causes them to detach from the coils and move towards the midplane of the vessel, where they merge together to form a single ST plasma. The relaxation from two flux-ropes with parallel currents to one ST plasma involves magnetic reconnection of poloidal field, as shown in Figure~\ref{fig:cartoon1}. 

With this technique, up to 0.5 MA of plasma current has been obtained, and electron and ion temperatures up to $1.2$ keV have been achieved on a timescale of $\approx 10$ ms\cite{ono2012mast,yamadaEPS} due to the high power reconnection heating.

Magnetic reconnection~\cite{biskamp00,priestbirn07} is a key-driver of particle acceleration and plasma heating in solar flares, the solar wind and the Earth magnetosphere. It is also thought to be an important process for topological relaxation of magnetic fields in pulsars, the solar dynamo and laboratory plasma devices\cite{zweibel09}. There are a number of dedicated magnetic reconnection experiments currently under operation~\cite{ono93,myamada97,egedal03,brown99,furno03} and progress has been made in validating theoretical models against laboratory data. Numerical simulations have been performed to model these experiments using fluid~\cite{lukinqin01,lukin03,gray10,myers11,murphy08} and Particle-In-Cell models~\cite{dorfman08}. These simulations include the coupling of the reconnection layer to the inductive coils, or to an ideal instability that drives the reconnection. Some also include the effects of the specific vessel geometry on the merging. It is clear from these studies that both of these effects are necessary to properly model the reconnection physics and to help interpret the experimental data. 

The merging-compression start-up method in MAST has been studied as a magnetic reconnection experiment~\cite{ono2012mast}. The strong magnetic fields ($\sim 0.5$ T) and low initial temperatures and densities $T_{e0}\approx T_{i0} \approx 10$ eV, $n\approx 5 \times 10^{18}$ m$^{-3}$ make this the highest Lundquist number and lowest plasma-$\beta$ reconnection experiment currently in operation~\cite{ono2012mast}. The results are therefore potentially relevant to astrophysical reconnection scenarios.   

In this paper we present results from non-linear two-fluid simulations of merging-compression start-up in MAST. In Section~\ref{model} we describe the two-fluid model and numerical methods used in this paper. Section~\ref{cartesian} describes the results from resistive MHD and Hall-MHD simulations of the merging in two-dimensional Cartesian geometry. In Section~\ref{toroidal} we describe the effect of the tight aspect ratio toroidal axisymmetric geometry on the merging for both the resistive and Hall-MHD cases. In Section~\ref{summary} we summarise the results presented.  

\section{\label{model}Two-fluid model}
\subsection{Fluid Equations}

Merging-compression occurs in the start-up phase of a MAST shot, when the plasma is colder and less dense than in a typical MAST flat-top phase. Our initial conditions are for a deuterium plasma with typical pre-merging values for the temperature $T_0 = T_{e0} = T_{i0} = 1.2 \times 10^5$ K ($= 10$ eV), density $n_0 = 5\times 10^{18}$ m$^{-3}$, magnetic field based on a typical toroidal field $B_0 = B_{T0} = 0.5$ T, and poloidal field of $B_{p0} = 0.1$ T. We take the typical length scale $L_0 = 1$ m, the order of the major and minor radii. With these values the toroidal Alfv\'en speed is $v_0 = 3.5 \times 10^6 \, \textrm{m \,s}^{-1}$ and $\tau_0  = L_0/v_0 = 0.29$ $\mu$s. Table~\ref{tab:params} shows normalised plasma parameters at merging-compression start-up.

\begin{table}
\caption{\label{tab:params} Characteristic plasma parameters calculated from $n_0$, $T_0$, $L_0$, $B_{T0}$, $B_{p0}$. These are the normalised resistivity $\eta$, the plasma beta calculated with the toroidal and poloidal fields $\beta_{T0}$ and $\beta_{p0}$, the ion and electron skin-depths $d_i$ and $d_e$, the ion and electron Larmor radii $\rho_i$ and $\rho_e$, the ion-sound radius $\rho_{is}$, the normalised ion viscosity $\mu$, and the normalised parallel and perpendicular heat conductivities $\kappa_\parallel$ and $\kappa_\perp$. }
\begin{ruledtabular}
\begin{tabular}{lcc}
Quantity & Value & Definition\\
\hline \\
$\eta$ & $10^{-5}$ & $\eta_{Sp,\parallel}/(\mu_0 v_0 L_0)$\footnotemark[1]\\
$\beta_{T0}$ & $8 \times 10^{-5}$ & $2\mu_0n_0k_BT_0/B_{T0}^2$ \\
$\beta_{p0}$ & $2 \times 10^{-3}$ & $2\mu_0n_0k_BT_0/B_{p0}^2$ \\
$d_i$ & $0.145$ & $c(n_0e^2/\epsilon_0m_i)^{-1/2}L_0^{-1}$ \\
$d_e$ & $2.4 \times 10^{-3}$ & $c(n_0e^2/\epsilon_0m_e)^{-1/2}L_0^{-1}$\\
$\rho_i$, $\rho_{is,0}$ & $9.3 \times 10^{-4}$ & $\sqrt{m_ik_BT_0}/(eB_0L_0)$\footnotemark[2] \\
$\rho_e$ & $1.5\times 10^{-5}$ &  $\sqrt{m_ek_BT_0}/(eB_0L_0)$ \\
$\mu$ & $10^{-3}$ & $\mu_i^\parallel/(m_in_0v_0L_0)=1/Re$\\
$\kappa_\parallel$ & $10^{-1}$ & $\kappa_e^\parallel/(L_0 v_0 n_0)$ \\
$\kappa_\perp$ & $10^{-7}$ & $\kappa_i^\perp/(L_0 v_0 n_0)$ \\

\end{tabular}
\end{ruledtabular}
\footnotetext[1]{We use an isotropic resistivity, $\eta$, calculated from the parallel Spitzer value $\eta_{Sp,\parallel}$ using the initial $n_0$ and $T_0$.}
\footnotetext[2]{Here $\rho_{is,0}$ is the ion-sound radius based on the initial electron temperature, $T_{e0}=T_0$, and typical magnetic field $B_0$.}
\end{table}

The fastest timescale at which the flux-ropes can merge is the ideal timescale of the attraction due to the parallel toroidal current. This is related to the poloidal Alfven time, $\tau_{A,p0} = (B_0/B_{p0}) \tau_0 \approx 1.5 \times 10^{-6}$ s. The ion and electron collision times are $\tau_{i,0} \approx 7 \times 10^{-6}$ s and $\tau_{e,0} \approx 10^{-7}$ s respectively. 
The electrons are collisional at this temperature and strongly magnetised ($\Omega_{ce,0} \tau_{e,0} \sim 10^4 \gg 1$ where $\Omega_{ce,0}$ is the initial electron gyro-frequency), so we model them with fluid equations\cite{braginskii}. The ions are also strongly magnetised ($\Omega_{ci,0}\tau_{i,0} \sim 10^2$) but semi-collisional. In this paper the ions are also treated as a fluid, and the possible implications of the departure from classical transport on these results will be left for future work. 

The Hall term is included within this fluid model, as the large ion skin-depth, $d_i = 14.5\,\textrm{cm}$, suggests that the decoupling of the ion and electron fluids due to ion inertia may be significant. The ion Larmor radius, $\rho_i$, and electron skin depth, $d_e$, are much smaller, although they can become comparable to the thinnest current sheets within these simulations. We leave the investigation of ion Finite Larmor Radius (FLR) effects and electron inertia for a future study. We do include the effects due to finite ion-sound radius, $\rho_{is}=\sqrt{T_e/m_i}/\Omega_{ci} = \rho_{is,0}\sqrt{T_e/T_0}B_0/B$, as previous studies have shown they are important in two-fluid reconnection with a strong guide field~\cite{kleva95,simakov10,schmidt09}. These studies often use a reduced model, based partly on the assumptions of large aspect ratio and uniform electron temperature. However, the MAST vessel is tight-aspect ratio and a large change in electron temperature is measured with the Thomson scattering diagnostic during the merging. We proceed in this paper with a fully compressible Hall-MHD fluid model including the scalar electron pressure term within Ohm's law~\cite{biskamp00}.

The governing equations are normalised by writing each dimensional variable $\bar{\chi}$ as $\bar{\chi} = \chi_0 \chi$, where $\chi_0$ is a typical dimensional constant for that variable and $\chi$ is the normalised variable. Choosing for example $E_0 = v_0 B_0$, $p_0 = B_0^2/\mu_0$ gives the normalised Hall-MHD equations

\begin{equation}\label{mass}\partial_t n + \boldsymbol{\nabla} \cdot (n\boldsymbol{v}_i) = 0,\end{equation}
\begin{equation}\label{mom}\partial_t (n \boldsymbol{v}_i) + \boldsymbol{\nabla} \cdot (n \boldsymbol{v}_i \boldsymbol{v}_i + p\mathbb{I} + \boldsymbol{\pi}_i) = \boldsymbol{j} \times \boldsymbol{B},\end{equation}
\begin{equation}\label{ohm}\boldsymbol{E} = -\boldsymbol{v}_e\times \boldsymbol{B} - \frac{d_i}{n}\boldsymbol{\nabla}p_e + \eta \boldsymbol{j} - \eta_H \nabla^2 \boldsymbol{j},\end{equation}
\begin{equation}\label{faraday}\partial_t \boldsymbol{B} = - \boldsymbol{\nabla} \times \boldsymbol{E},\end{equation}
\begin{eqnarray}\label{pressure}&(\gamma - 1)^{-1}[\partial_t p + \boldsymbol{v}_i \cdot \boldsymbol{\nabla}p + \gamma p\boldsymbol{\nabla}\cdot \boldsymbol{v}_i] =\\ \nonumber
&\eta j^2 + \eta_H (\boldsymbol{\nabla}\boldsymbol{j})^2 - \boldsymbol{\pi}_i:\boldsymbol{\nabla}\boldsymbol{v}_i - \boldsymbol{\nabla} \cdot \boldsymbol{q}.\end{eqnarray}

Here, $n$ is the plasma density, $\boldsymbol{v}_i$ the ion velocity, $\boldsymbol{v}_e = \boldsymbol{v}_i - d_i \boldsymbol{j}/n$ is the electron velocity where $d_i$ is the normalised ion skin-depth (Table~\ref{tab:params}), $\boldsymbol{B}$ the magnetic field, $p=p_i+p_e$ the total (sum of ion and electron) thermal pressures where we assume $p_i = p_e = p/2$, $\mathbb{I}$ is the unit tensor, $\boldsymbol{j} = \boldsymbol{\nabla}\times \boldsymbol{B}$ is the current density, and $\boldsymbol{E}$ the electric field. The ion stress tensor is $\boldsymbol{\pi}_i = -\mu(\boldsymbol{\nabla}\boldsymbol{v}_i + \boldsymbol{\nabla}\boldsymbol{v}_i^T)$, and the heat-flux vector $\boldsymbol{q}$ has the simplified anisotropic form $\boldsymbol{q} = -\kappa_\parallel \boldsymbol{\nabla}_\parallel T - \kappa_\perp \boldsymbol{\nabla} T$ where $\boldsymbol{\nabla}_\parallel = \boldsymbol{\hat{b}}(\boldsymbol{\hat{b}}\cdot \boldsymbol{\nabla})$. 

The coefficients in equations~(\ref{mass}-\ref{pressure}) are the normalised resistivity $\eta$, the hyper-resistivity $\eta_H$ (see below), the ratio of specific heats $\gamma=5/3$, the normalised ion viscosity $\mu$, and the normalised parallel and perpendicular heat conductivities $\kappa_\parallel$ and $\kappa_\perp$. These coefficients are taken to be constant and uniform with the values in Table~\ref{tab:params} for all results unless explicitly stated otherwise. Note the ion viscosity $\mu$ is based on the initial parallel value $\mu_i^\parallel$, rather than the perpendicular value, for numerical stability. The viscosity is therefore treated as a free parameter in the model, and is varied in the simulations described below.

The final term in equation~(\ref{ohm}) is a hyper-resistive current diffusion term~\cite{biskamp00}, that is common in Hall-MHD simulations of reconnection (see \citet{bhattachar01gem} and references therein). This term is used to set a dissipation scale for waves that have a quadratic dispersion relation. It can also give parallel electric field at the X-point ($\boldsymbol{\hat{b}}\cdot \boldsymbol{E}$), and so contributes to breaking the frozen-in condition. Physically this term is related to an electron viscosity inside the current sheet where the electrons carry most of the current density (for $|\boldsymbol{v}_e| \gg |\boldsymbol{v}_i|$ the term $-\nabla^2 \boldsymbol{j} \approx \nabla^2 n\boldsymbol{v}_e/d_i$). The value of $\eta_H$ is taken as a free parameter in this model. We also include the associated hyper-resistive heating term in equation~(\ref{pressure}), which has often been neglected in other studies but it is required to satisfy energy conservation. 

\subsection{Code}

The equations~(\ref{mass}-\ref{pressure}) are solved using the two-dimensional implementation of the high order finite (spectral) element framework HiFi\cite{glassertang04,lukinthesis}. The problem is discretised on a grid of $N_R \times N_Z$ finite-elements, with each element having order $N_p$ Jacobi polynomial basis functions. The grid is stretched in both $R$ and $Z$ directions to give high resolution in regions of interest, namely the current sheet. The effective resolution of this scheme is $(N_R\times N_p)\times (N_Z\times N_p)$. We use $N_p=4$ in all results presented here. The values of $N_R$ and $N_Z$ vary depending on the magnitude of the dissipation coefficients used, and are stated explicitly for each simulation. For simulations with the lowest dissipation scales we perform grid convergence tests by coarsening $N_R$ and $N_Z$ by a factor of $2$.

All of the simulations described in this paper use the implicit Crank-Nicolson method for time advance, to avoid the need for a prohibitively short time-step in the presence of dispersive waves. However, the maximum timestep is limited to ensure accuracy. For resistive MHD simulations ($d_i=0$) the maximum timestep is limited to $\Delta t = 5\times 10^{-2} \tau_0$, and the maximum timestep used for Hall-MHD simulations is $\Delta t = 10^{-2}\tau_0$. Note that the actual timestep is adaptively determined by the code with respect to given convergence criteria and is often much shorter than this.

\subsection{Initial Conditions}

The initial conditions are two flux-ropes with parallel toroidal current and strong toroidal (guide) field inside the MAST vacuum vessel. The domain is $R\in [0.2,2.0\,\textrm{m}]$, $Z \in [-2.2,2.2\,\textrm{m}]$, where the inner radial value is the radius of the centre post, and the other values specify the outer walls of the vessel. We do not model the in-vessel poloidal field coils, and thus the complicated physics of breakdown and flux-rope formation. These initial conditions correspond to the time after the flux-ropes have detached from the P3 coils, but before they have moved towards the mid-plane (note that this detachment prior to merging is inferred from fast camera images, see eg. Figure~1c in \citet{yamadaEPS}).

The standard equilibrium fitting routines\cite{lao90} cannot be used when there are two separate sets of nested flux-surfaces in the plasma, and so at present it is not possible to reconstruct the magnetic structure of the flux-ropes prior to merging. However, the total current in the domain, $I_{\textrm{plasma}}$, is known and the width of the flux-ropes is estimated from the fast camera images. For the Cartesian simulations, $(R,T,Z)$ where $T$ is the invariant out-of-plane (infinite aspect-ratio toroidal) direction, we construct each of the flux-ropes using the smooth 1D current profile

\[
j_T(r) =
\begin{cases}
  j_m \left(1-\left(r/w\right)^2\right)^2 & \text{if } r \leq w, \\
  0 & \text{if } r > w,
\end{cases}
\]
where $j_T$ is the out-of-plane current density, $r = \sqrt{(R-R_0)^2 + (Z\pm Z_0)^2}$ is the radial distance from the centre of each flux rope ($R=R_0$, $Z=\pm Z_0$), $w$ is the flux-rope radius, and $j_m$ is the maximum current density. In all of the simulations presented we use $w=0.4$ m and $j_m = 2 \, [B_0/(\mu_0L_0)] = 0.8 \, \textrm{MA\,m}^{-2}$ to give total current $I_{\textrm{plasma}} = 2\times(\pi j_m w^2/3)=268 \,\textrm{kA}$, the same as MAST shot 25740 for which experimental results are presented in~\citet{ono2012mast}. 

\begin{figure}
\includegraphics[width=0.48\textwidth]{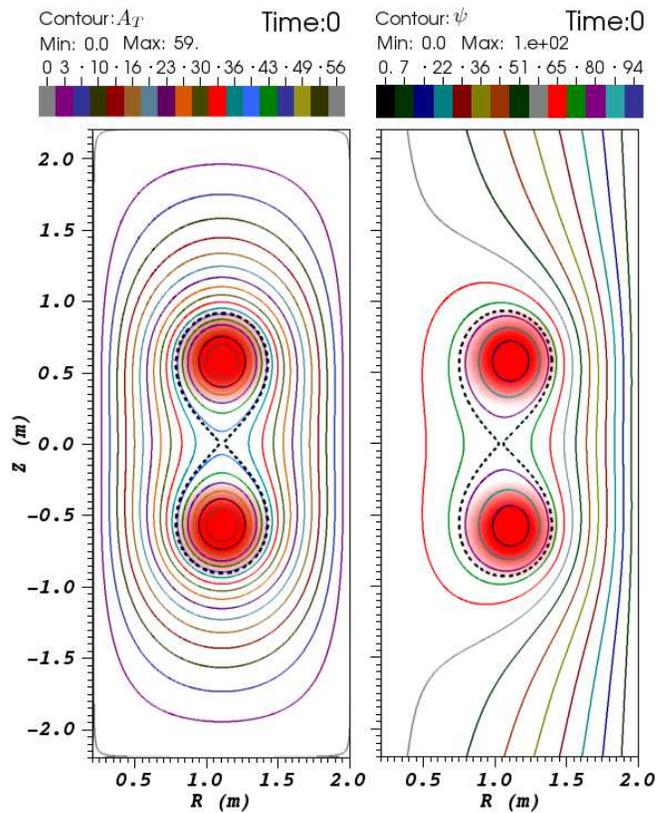}
\caption{\label{fig:init} The initial conditions for Cartesian simulations (left) and toroidal axisymmetric simulations (right) for initial half-separation $a=0.6$ m, flux-rope radius $w=0.4$ m, and maximum current density $j_m = 8\times 10^5 \, \textrm{A \,m}^{-2}$. The colour scale is the current density $j_T$ ($j_\phi$) and the coloured contours are contours of $A_T$ ($\psi = R A_\phi$) for the Cartesian (toroidal) simulations. The black dashed lines correspond to $A_{T,Xpt} = 38.9 \, \textrm{mWb \, m}^{-1}$, and $\psi_{Xpt} = 76.5 \, \textrm{mWb}$ for the Cartesian and toroidal simulations respectively. }
\end{figure}

Due to the low poloidal beta $\beta_{p0} = 2 \times 10^{-3}$, the internal pinch-force of each flux-rope is balanced by a paramagnetic increase in $B_T$. This kind of magnetic profile has been measured in ideally relaxing flux-ropes under a strong toroidal field in the TS-3 merging device~\cite{ono2012mast}. Firstly, the poloidal magnetic field due to the 1D current profile was found, matching the outer potential solution to the solution inside the flux-rope at $r=w$. This was then used in the 1-D force-free equilibrium equation to find the required $B_T$ for radial force balance
\[
B_T =
\begin{cases}-j_m\Big(\frac{B_{T0}^2}{j_m^2} + \frac{47w^2}{360} - \frac{r^2}{2} + \frac{3r^4}{4w^2} \\
 - \frac{5r^6}{9w^4} + \frac{5r^8}{24w^6} - \frac{r^{10}}{30w^8}\Big)^{1/2} & \hspace{-0.2 em} \text{if } r \leq w\\
-B_{T0} & \hspace{-0.2 em}\text{if } r > w
\end{cases}
\]
where the sign is determined by the respective orientations of the toroidal field and plasma current in MAST (the toroidal field is clock-wise as viewed from above). The out-of-plane magnetic potential is found by solving $-\nabla^2 A_T = j_T$, subject to the boundary condition $A_T = 0$, with the HiFi framework. With this method each flux-rope is very close to force-free with respect to the internal pinch force, but there is finite Lorentz force between the flux-ropes that causes them to mutually attract. A large initial separation, $2a = 1.2$ m, is chosen so that this force is small at $t=0$. The initial conditions for the Cartesian simulations are shown in the left hand panel of Figure~\ref{fig:init}. A two-dimensional separatrix, with an X-point, separates the ``public flux'' contours that enclose both flux-ropes without breaking, and the ``private flux'' of each flux-rope which is available for reconnecction.   

The initial conditions for the 2D toroidal axisymmetric simulations $(R,\phi,Z)$ are set up in the same manner, but the toroidal magnetic potential $A_\phi$ is solved subject to the boundary conditions $A_\phi = B_V R/2$. This gives a uniform vertical field of magnitude $B_V$ that reduces the radially outwards hoop force on the flux-ropes. We use $B_v = -0.06 B_0 = -0.03$ T unless specified otherwise. In toroidal geometry, the vacuum toroidal field is $B_{\phi0} = -B_{T0} R_0/R$ where $R_0 = 0.85$ m is the major radius. 

After the initial conditions are set-up, the equations~(\ref{mass}-\ref{pressure}) are advanced in Cartesian geometry with conducting wall boundary conditions $\partial_t A_T = 0$, zero tangential current $\boldsymbol{\hat{n}}\cdot \boldsymbol{\nabla} B_T = 0$, $j_T = 0$, perfect slip solid wall $\boldsymbol{\hat{n}}\cdot \boldsymbol{\nabla} (\boldsymbol{\hat{n}}\times \boldsymbol{v}_i) = \boldsymbol{0}$ and $\boldsymbol{\hat{n}}\cdot \boldsymbol{v}_i = 0$, and no temperature gradient $\boldsymbol{\hat{n}}\cdot \boldsymbol{\nabla} T = 0$. In toroidal geometry, where there is a normal component of the field intersecting the vertical boundaries, the stricter condition $\boldsymbol{v}_i=0$ is used to ensure there is no tangential convective electric field and associated normal Poynting flux through the boundary.

\begin{figure*}
\centering
\includegraphics[width=0.85\textwidth]{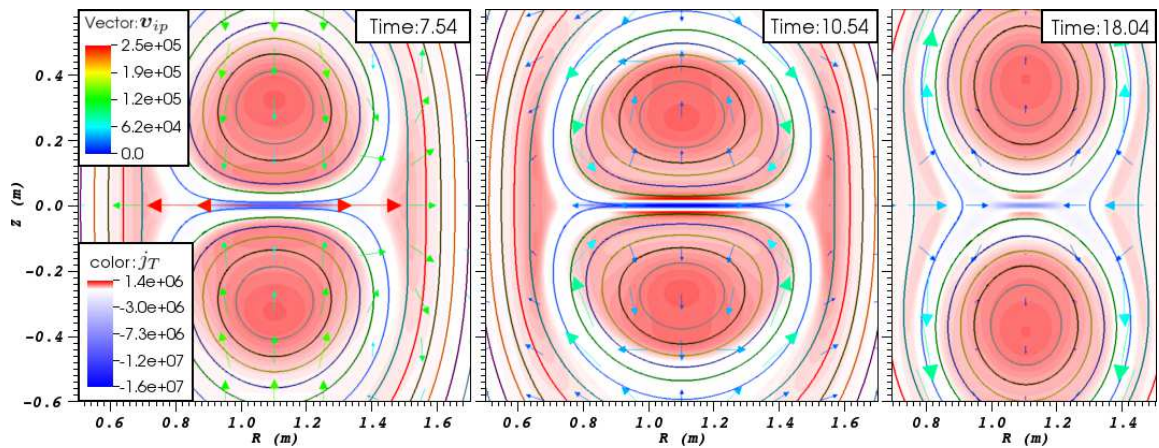}
\caption{\label{fig:slosh}Snapshots of the flux-ropes at $t=7.54 \tau_0$, $t=10.54 \tau_0$, and $t=18.04 \tau_0$ (where $\tau_0 = 0.29$ $\mu$s) for a resistive MHD simulation ($d_i=\eta_H=0$) with $\eta=10^{-5}$ and $\mu=10^{-3}$. The out-of-plane current density, $j_T$ in A m$^{-2}$, is shown in colour, the coloured lines are contours of the out-of-plane potential, $A_T$, and the coloured vectors are in-plane velocity vectors, $\boldsymbol{v}_{ip}$ in $\textrm{m\, s}^{-1}$.}
\end{figure*}

\section{\label{cartesian}Resistive and Hall-MHD simulations in Cartesian geometry}

\subsection{Resistive MHD}

The first set of simulations described here are resistive MHD simulations in Cartesian $(R,T,Z)$ geometry (the equations~(\ref{mass}-\ref{pressure}) were solved with $d_i=\eta_H =0$). All the resistive MHD simulations use $N_R = 180$, $N_Z = 360$ giving an effective resolution of $720\times 1440$ ($N_p = 4$). The grid is stretched so that the minimum grid spacing in the $Z$ direction is $\Delta Z = 2.3 \times 10^{-4}$ m at the midplane $Z=0$. The initial conditions are shown in the left-hand panel of Figure~\ref{fig:init}. There is $A_{T,\textrm{pr}} = A_{T,\textrm{max}} - A_{T,\textrm{Xpt}} = 59 - 38.9 = 20.1 \, \textrm{mWb \,m}^{-1}$ of private flux within each flux-rope at the start of the simulation. 

Initially the thermal pressure is uniform but a finite Lorentz force between the flux-ropes, due to their parallel toroidal currents, causes them to be mutually attracted and they move towards the midplane. As the flux-ropes move together, a build up in the out-of-plane magnetic field $B_T$ in the region between them reduces the initial acceleration.  

Figure~\ref{fig:slosh} shows the out-of-plane current density, $j_T$, the in-plane plasma velocity, $\boldsymbol{v}_{ip}=(v_{iR},v_{iZ})$, and the out-of-plane magnetic potential, $A_T$, at three snapshots during the merging. At $t = 7.54 \, \tau_0 = 2.2 \, \mu s$ the X-point between the two flux-ropes collapses forming a thin sheet of negative out-of-plane current (blue). The plasma between the flux-ropes is accelerated in the radial direction in two jets, reaching a maximum velocity of $v_{R,max} \approx 2.5 \times 10^5 \,\textrm{m\,s}^{-1}$.

At $t= 10.54 \,\tau_0 = 3.06 \,\mu s$ (middle panel) the outflow speed drops to $v_{R} = 5\times 10^4 \, \textrm{m\,s}^{-1}$ (small blue outflow arrows), and the vortical plasma flows that bring in flux to the current sheet have reversed direction. The flattened front edges of the flux-ropes indicates a strong flux pile-up of the reconnecting field, $B_R$, resulting in two layers of positive current (red) on the vertical edges of the current sheet. The repulsive Lorentz force between the negative current sheet and these oppositely directed current layers can prevent plasma from entering the sheet.

 At $t=18.04 \, \tau_0 = 5.23 \, \mu s$ the O-points at the centre of the two flux-ropes have clearly moved apart, similar to the sloshing motion studied in simulations of the coalescence instability~\cite{biskampwelt80}. The value of $\eta$ used in this simulation is similar to that of \citet{knollchacon06}, who find this reversed O-point motion for $\eta \le 2 \times 10^{-5}$. However, \citet{knollchacon06} set $\mu=\eta$, whereas we have $\mu > \eta$ for this simulation. During this sloshing the area inside the outer-most private flux contour (green contour) changes by less than $1\%$ between the first and third panels. 

\begin{figure}
\centering
\includegraphics[width=0.46\textwidth]{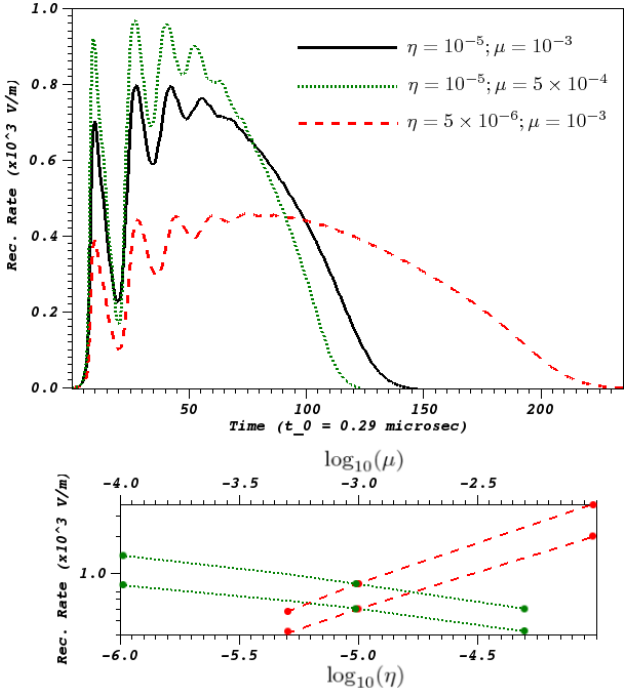}
\caption{\label{fig:sloshrate}Above: The reconnection rate, $\partial_t A_T$ in $\textrm{V\,m}^{-1}$, plotted over time for the Cartesian resistive-MHD simulation with $\eta=10^{-5}$, $\mu=10^{-3}$ black-solid line; $\eta=10^{-5}$, $\mu=5\times 10^{-4}$ green-dotted line; $\eta=5\times 10^{-6}$, $\mu=10^{-3}$ the red-dashed line. Below: The peak, $(\partial_t A_T)_{\textrm{max}}$ (top pair of lines), and average, $<\partial_t A_T>$ (bottom pair of lines), reconnection rates. The green-dotted lines are for fixed $\eta=10^{-5}$ with various $\mu$ (top axis), and the red-dashed lines are for fixed $\mu=10^{-3}$ with various $\eta$ (bottom axis).} 
\end{figure}

The reconnection rate in Cartesian geometry is $\partial_t A_T = - E_T$ at the location of the X-point ($R=1.1$ m, $Z=0$ m). This reconnection rate for this simulation is plotted against time in the top panel of Figure~\ref{fig:sloshrate} (black-solid line). The repeated sloshing of the flux-ropes modulates the reconnection rate through large amplitude oscillations with a period of a few poloidal Alfv\'en times, $\approx 4 \tau_p$ where $\tau_p \approx 5 \tau_0$. This is followed by a gradual decaying phase as the flux-ropes shrink, until all of the available flux is reconnected. The global maximum reconnection rate is $800 \, \textrm{V \,m}^{-1}$ at $t=27.54 \, \tau_0 = 7.99 \,\mu s$ (second peak), when the Full-Width Half-Minimum (FWHM) length of the current sheet is $\Delta_{FWHM} \approx 0.32$ m and the width is $\delta_{FWHM} \approx 1.1 \times 10^{-2}$ m. An aspect ratio of $30$ is consistent with the visco-resistive scaling for the Sweet-Parker sheet~\cite{park84} $\delta_{\mu \eta} \sim \eta_{eff}^{1/4}\,\mu_{eff}^{1/4} \approx \Delta/29.6$, where $\eta_{eff} \equiv S^{-1} = \eta\, v_0\, L_0/(v_{A,in}\, \Delta)$ and $\mu_{eff} = \mu\, v_0 \, L_0/(v_{A,in} \,\Delta)$ are the effective inverse Lundquist number and inverse Reynolds number respectively, defined in terms of the current sheet length $\Delta=\Delta_{FWHM}$, and the Alfv\'en velocity due to the reconnecting component of the field at the sheet edge $v_{A,in}=(B_{R,in}/B_0)\sqrt{n_0/n_{in}}\,v_0 = 0.2742 v_0$.

The top panel of Figure~\ref{fig:sloshrate} also shows the reconnection rate against time for two other simulations. When the viscosity is reduced by a factor of two ($\eta = 10^{-5}$, $\mu=5\times 10^{-4}$ green-dotted), the peak reconnection rate increases and the total merge time decreases from $T_{merge} = 142.7\, \tau_0 = 41.4 \, \mu s$ to $122 \, \tau_0 = 35.4 \, \mu s$. A factor-of-two reduction in resistivity ($\eta=5\times 10^{-6}$, $\mu=10^{-3}$ red-dashed line) increases the merge time to $231 \, \tau_0 = 67 \, \mu s$. As the total flux reconnected is the same in each simulation $A_{T,\textrm{pr}} = 20.1 \textrm{mWb \,m}^{-1}$ the average reconnection rate can be calculated as $<\partial_t A_T> = A_{T,\textrm{pr}}/T_{merge}$, so for the case of $\eta=10^{-5}$, $\mu=10^{-3}$ the average reconnection rate is $<\partial_t A_T> = 486 \, \textrm{V \, m}^{-1}$. The bottom panel of Figure~\ref{fig:sloshrate} shows the scalings for the peak $(\partial_t A_T)_\textrm{max}$ and average $<\partial_t A_T>$ reconnection rates, in $\textrm{V \, m}^{-1}$, against resistivity (red-dashed line, bottom axis) and against viscosity (green-dotted line, top axis). The peak reconnection rates (note the maximum value is not always on the second bounce) scale as $(\partial_t A_T)_\textrm{max} \sim \eta^{0.69} \mu^{-0.26}$ and the average reconnection rates as $<\partial_t A_T> \sim \eta^{0.62} \mu^{-0.23}$. These are in good agreement with both the visco-resistive scalings~\cite{park84} for a Sweet-Parker current sheet, $\sim \eta^{1/2} (1 + \mu/\eta)^{-1/4}$, and a previous study\cite{breslaujardin03} of coalescing flux-ropes with large magnetic Prandtl number $Pr_m=\mu/\eta$, who find $<\partial_t A_T> \sim \eta^{0.6}\mu^{-0.3}$. Although, the latter study does not mention the sloshing effect, and gives scalings only for the average reconnection rate.

\begin{figure}
\centering
\includegraphics[width=0.44\textwidth]{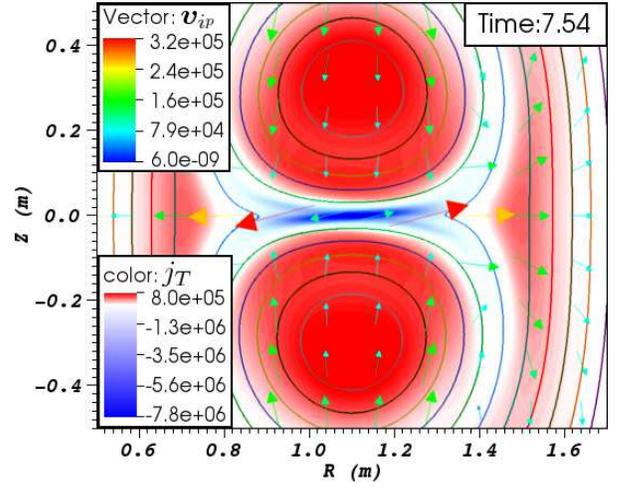}
\caption{\label{fig:halltilt}Snapshot of the Cartesian Hall-MHD simulation with $\eta_H=10^{-8}$, $d_i=0.145$. The current density, $j_T$ in $\textrm{A\,m}^{-2}$, is shown in colour and the magnetic potential $A_T$ is shown as coloured contours. Coloured arrows show the ion in-plane velocity, $\boldsymbol{v}_{ip}$ in $\textrm{m\,s}^{-1}$, where the middle of the arrow is the point at which the velocity field is sampled.}
\end{figure}

\subsection{Hall-MHD}

The effects of the Hall and hyper-resistive terms on the merging are reported here. The standard Hall-MHD simulation has the same parameters and numerical grid as the standard resistive MHD simulation ($\eta=10^{-5}$, $\mu=10^{-3}$) but with $d_i = 0.145$ m and $\eta_H = 10^{-8}$. A snapshot of the current density, $j_T$, and the in-plane ion velocity, $\boldsymbol{v}_{ip}$, at $t=7.54 \, \tau_0 = 2.2 \, \mu s$ is shown in Figure~\ref{fig:halltilt}. There are several differences evident when comparing this figure with the previous simulation in Figure~\ref{fig:slosh}. Quantitatively, the time average reconnection rate up to this snapshot is $361 \, \textrm{V\,m}^{-1}$ compared to $50 \, \textrm{V\,m}^{-1}$ for the resistive simulation (the amount of flux reconnected at this time in the Hall-MHD simulation is roughly the same as that in the third panel of Figure~\ref{fig:slosh}). However, this is partly due to the increase in dissipation scale when including hyper-resistivity, as well as normal resistivity, to break the frozen in condition (see below for scalings with $\eta_H$). Qualitatively, there is a clear tilt of the ion outflow jets when the Hall term is switched on (this tilt is not present for $d_i = 0$ with $\eta_H \neq 0$). The outflow jets have a positive (negative) vertical component on the outer (inner) radial side. There is also a tilt of the main current sheet, as the current density is stronger across the bottom separator on the inner radial side, and the top separator on the outer side. The radial length of the current sheet measured at $Z=0$ at this time is $L \approx 20$ cm, which is of the same order as the initial flux-rope radius ($w=0.4$ m).

\begin{figure}
\centering
\includegraphics[width=0.5\textwidth]{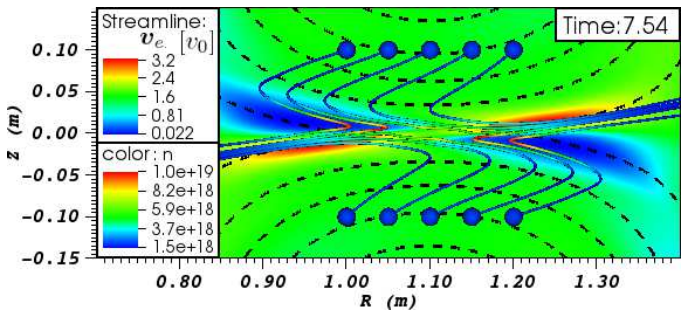}
\caption{\label{fig:denscart}Number density, $n$ in $m^{-3}$, in colour scale and bulk electron velocity, $\boldsymbol{v}_e$ in units of $v_0 = 3.5\times 10^{6} \,\textrm{m\,s}^{-1}$, streamlines (coloured tubes which start at the locations of the blue circles) for the Hall-MHD simulation with $\eta_H=10^{-8}$. The black-dashed lines show contours of the magnetic potential $A_T$.}
\end{figure}

Figure~\ref{fig:denscart} shows the plasma number density, $n$, for the same simulation and at the same time as in Figure~\ref{fig:halltilt}. There are $\mathcal{O}(1)$ density variations in a quadrupole-like shape within the diffusion region between the two flux-ropes, similar to what has been found in previous fluid~\cite{kleva95,huba05} and Particle-In-Cell (PIC) simulations~\cite{ricci04}. Over-plotted are streamlines of bulk electron velocity $\boldsymbol{v}_e = \boldsymbol{v}_i - d_i \boldsymbol{j}/n$. Within the flux-ropes the motion of the electrons is dominated by perpendicular drifts towards the current sheet. However, within the diffusion region the streamlines become nearly parallel to the in-plane field. The electrons are accelerated in bulk within the density cavities and slow down as they enter the high-density regions. One study\cite{kleva95} has suggested that this density structure can be caused by parallel electron compressibility in strong guide field reconnection. We verified this by overplotting contours of $ \boldsymbol{\hat{b}}\cdot \boldsymbol{\nabla} (\boldsymbol{\hat{b}}\cdot \boldsymbol{v}_e)$ (not shown), finding that large positive (negative) values overlie the density cavities (peaks). Also, the magnitude of this term is in good agreement with the magnitude of $\boldsymbol{\nabla} \cdot \boldsymbol{v}_e$ at this time. The same study suggests that these features are localised within an ion-sound radius $\rho_{is} = \sqrt{T_e/m_i}/\Omega_{ci} = \sqrt{\beta_e/2} d_i$. In these simulations the value of $\rho_{is}$ varies in space and time, strongly depending on the balance between the heating and thermal conduction terms in the energy equation. At $t=7.54 \, \tau_0 = 2.2 \, \mu s$ $\rho_{is} \approx 1 \, \textrm{cm}$ within the current sheet. We also ran simulations with larger $B_{T0}$ finding the difference between the maximum and minimum density values is reduced, consistent with the suggestion that this strong guide-field two-fluid reconnection signature scales with $\rho_{is} \propto B^{-1}$.

\begin{figure}
\centering
\includegraphics[width=0.49\textwidth]{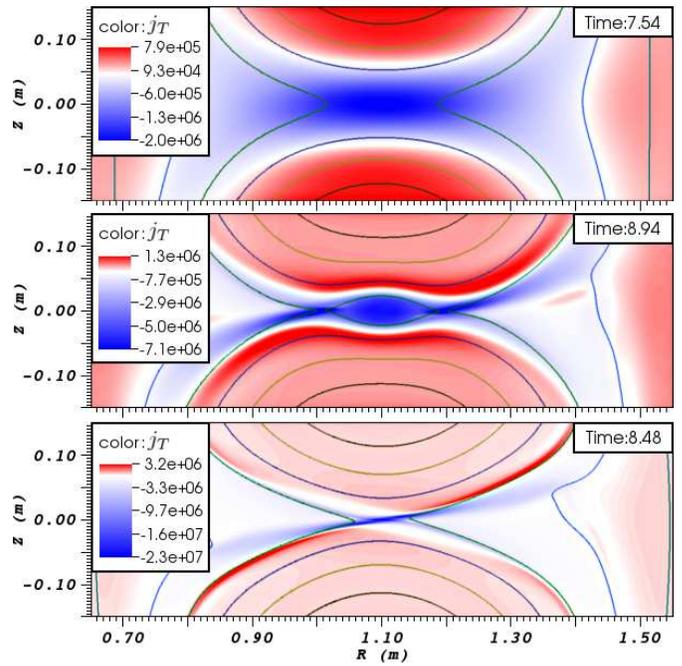}
\caption{\label{fig:latetime}The current density, $j_T$ in colour, and contours of the magnetic potential, $A_T$, for three Cartesian Hall-MHD simulations with $\eta_H = 10^{-6}$ (top), $\eta_H=10^{-8}$ (middle) and $\eta_H = 10^{-10}$ (bottom). For all three simulations $\eta = 10^{-5}$, $\mu=10^{-3}$ and $d_i=0.145$. }
\end{figure}

To examine how the strength of the dissipation effects the merging, we vary the hyper-resistivity $\eta_H$. We also ran simulations varying the resistivity $\eta$, but found that both the qualitative picture of the merging and the reconnection rate were insensitive to $\eta$. This is consistent with the hyper-resistive contribution to the reconnection electric field at the X-point dominating the resistive contribution in these simulations. Figure~\ref{fig:latetime} shows the out-of-plane current density, $j_T$, and contours of the out-of-plane potential, $A_T$, for three simulations, at a time when the same amount of flux has been reconnected (green contour), for $\eta_H = 10^{-6}$ (top), $\eta_H = 10^{-8}$ (middle) and $\eta_H = 10^{-10}$ (bottom). The simulation in the middle panel is the same one as in Figures~\ref{fig:halltilt} and~\ref{fig:denscart} but at a later time. The other two simulations have identical initial conditions and plasma parameters except for the hyper-resistivity. There are $N_R=180$, $N_Z=360$ finite elements for the simulation with $\eta_H = 10^{-6}$, and $N_R = 360$, $N_Z = 540$ for $\eta_H = 10^{-10}$, where a convergence test has been run on the latter (see below).

For $\eta_H = 10^{-8}$, the current sheet that was forming in Figure~\ref{fig:halltilt} has become unstable to a tearing-type instability, forming an island at the centre of the sheet before any local peak in the reconnection rate. A $180^\circ$ rotational symmetry that is present in the initial conditions is preserved by equations~(\ref{mass}-\ref{pressure}), so there is no preferred direction for the island to be ejected. The island grows as flux is reconnected at two x-points on either side of the island, and saturates when the internal magnetic pressure balances the attractive force between the flux-ropes. At saturation the reconnection stalls and the whole system oscillates as it relaxes.

\begin{figure}
\centering
\includegraphics[width=0.5\textwidth]{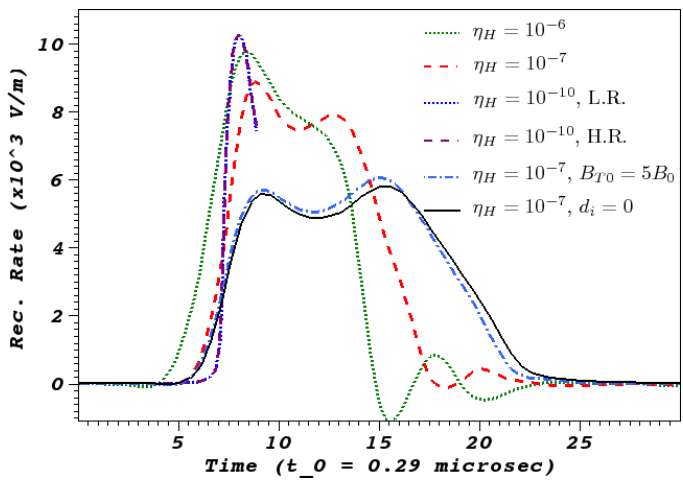}
\caption{\label{fig:hyp-rate}The reconnection rate, $\partial_t {A}_T$, plotted against time for standard Hall-MHD simulations ($d_i=0.145$, $B_{T,0} = 1B_0 = 0.5$ T) with hyper-resistivity $\eta_H=10^{-6}$ (green dotted), $\eta_H = 10^{-7}$ (red dashed), and $\eta_H = 10^{-10}$. The rate for the lowest hyper-resistivity is shown for two different resolutions: ``LR'' where $N_R = 180$ and $N_Z = 270$ (blue-dotted), and ``HR'' where $N_R = 360$ and $N_Z = 540$ (purple dashed). Also plotted is a simulation with $d_i=0.145$ m, $\eta_H=10^{-7}$ and strong toroidal field $B_T = 5 B_0 = 2.5$ T, and a hyper-resistive MHD simulation $d_i = 0$, $\eta_H = 10^{-7}$. }
\end{figure}

For the case of $\eta_H=10^{-6}$, the width and length of the current sheet are $\delta_{FWHM} = 7.3$ cm and $\Delta_{FWHM} = 33$ cm respectively (note that this is slightly before the peak in the reconnection rate, see Figure~\ref{fig:hyp-rate}). The ion-sound radius is $\rho_{is} \approx 1$ cm within the sheet at this time. The small-aspect-ratio current sheet is stable against break-up for the duration of the merging. 
For the simulation with $\eta_H = 10^{-10}$ there is a localised region of intense current at the X-point with length $\Delta_{FWHM} = 4.4$ cm measured at $Z=0$, and width $\delta_{FWHM} = 6.5$ mm compared to $\rho_{is} = 3.1$ cm measured at the X-point at this time. Note that the strong current gives localised hyper-resistive heating, increasing the temperature despite reductions in $\eta_H$. The separatrices of the X-point have opened up in the outflow region, and there are sharp gradients in the current density across the separatrices consistent with classical pictures of fast reconnection~\cite{petschek64}. A threshold current sheet width of the ion sound radius, $\rho_{is}$, for fast-reconnection is in agreement with previous studies of strong guide-field reconnection~\cite{kleva95,simakov10,schmidt09}. After this snapshot the separatrices are pushed closed again and there are off-centre island pairs formed and ejected from the current sheet.

The time of the snapshot in the bottom panel ($\eta_H=10^{-10}$) is later than that for the top-panel, suggesting a slower average reconnection rate. To investigate this, we plot the reconnection rate against time for three values of the hyper-resistivity, $\eta_H=10^{-6}$, $10^{-7}$ and $10^{-10}$, in Figure~\ref{fig:hyp-rate}. Note that for $\eta_H = 10^{-8}$, $10^{-9}$ the sheet becomes unstable before any peak in the reconnection rate. Both the peak and the average reconnection rates decrease as $\eta_H$ is reduced from $10^{-6}$ to $10^{-7}$, and there is greater oscillation in the case with lower dissipation suggesting that the merging is approaching a sloshing-regime. The simulation with $\eta_H = 10^{-10}$ has much slower reconnection rate initially, but increases at $t\approx 6.5 \, \tau_0 = 1.9 \, \mu s$ (the current sheet width $\delta_{FWHM}$ drops below $\rho_{is}$ at $t=7 \, \tau_0$, after which the outflow separatrices open up and the sheet is localised in the radial direction), to give a higher peak reconnection rate than in the simulations with larger $\eta_H$. The simulation with $\eta_H=10^{-10}$ was run at two different grid-resolutions to verify that the solution is converged. The relative change in the peak reconnection rate from $N_R = 180$, $N_Z = 270$ to $N_R = 360$, $N_Z = 540$ is only 0.2\%. These curves are only plotted over the time period which the current sheet is stable (before island formation). Also plotted in Figure~\ref{fig:hyp-rate} is the reconnection rate for a single-fluid ($d_i=0$) hyper-resistive MHD simulation with $\eta_H = 10^{-7}$ (black-solid line), and the reconnection rate for a Hall-MHD simulation with $d_i = 0.145$ m, $\eta_H = 10^{-7}$ with much stronger out-of-plane field $B_T = 5 B_0 = 2.5$ T. Increasing the out-of-plane field reduces the ion-sound radius, as $\rho_{is} \propto B_T^{-1}$, and suppresses the mechanism of fast reconnection. The reconnection rate tends towards the collisional limit (the rate for $d_i = \rho_{is} = 0$). 

\subsection{Discussion}

It is important to determine whether merging-compression start-up in MAST lies within the purely collisional or open X-point regimes. It was shown above that the open X-point regime occurs when the current sheet width, $\delta$, drops below the ion-sound radius, $\rho_{is}$. 

The value of $\rho_{is}$ can be estimated directly from the experimental data. In merging-compression experiments electron temperatures have been measured in the range $10 \, \textrm{eV} \lesssim T_e \lesssim 1000 \, \textrm{eV}$, which, along with a typical toroidal field strength of $0.5$ T, gives $\rho_{is} = 0.93 - 9.3$ mm. Within the series of resistive and Hall-MHD simulations listed above, some runs with the lowest dissipation coefficients have current sheet widths within this range (e.g. the Hall-MHD simulation with $\eta_H = 10^{-10}$, $\eta=10^{-6}$ and $\mu = 10^{-3}$, or the resistive MHD simulation with $\eta = 10^{-5}$, $\mu = 10^{-4}$ that has a minimum width of $\delta = 7.59$ mm at the peak reconnection rate). Also, for numerical reasons, the values of the ion viscosity and hyper-resistivity (electron viscosity) have been enhanced with respect to the perpendicular values. For instance, a normalised perpendicular ion-viscosity~\cite{braginskii} calculated using the initial $T_0$ and $n_0$ would be $\mu = 10^{-7}$. It seems likely that the current sheet width can drop below the ion-sound radius for realistic merging-compression values of the ion and electron viscosities. This open X-point configuration, with a radially localised current sheet, may explain a narrow electron temperature peak of ~2-3 cm width found in the experimental data~\cite{ono2012mast}, provided that the electron heating is co-spatial with current (e.g. Ohmic heating).   
 
\section{\label{toroidal}Effects of tight aspect-ratio toroidal axisymmetric geometry}

\begin{figure}
\centering
\includegraphics[width=0.45\textwidth]{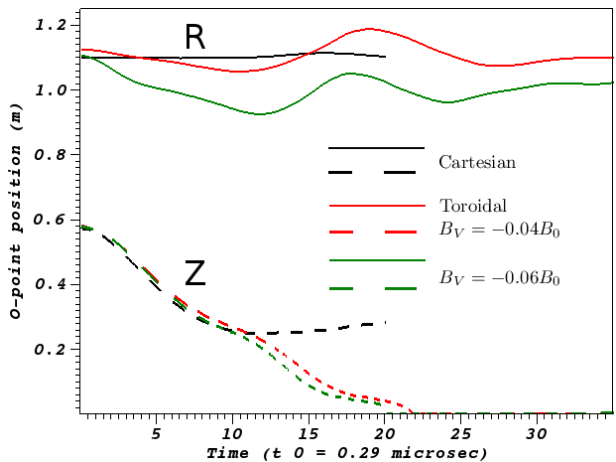}
\caption{\label{fig:opoints}The $R$ (top, solid lines) and $Z$ (bottom, dashed lines) positions of the O-points for the standard Cartesian Hall-MHD simulation (black), and the standard toroidal axisymmetric Hall-MHD simulation with vertical flux of $B_V=-0.04\,B_0 = -0.02 \,\textrm{T}$ (red) and $B_V = -0.06 \, B_0 = -0.03 \,\textrm{T}$ (green).}
\end{figure}

Figure~\ref{fig:opoints} shows the radial, $R$, and vertical, $Z$, positions of the flux-rope O-points over time for three simulations. One is the standard Cartesian Hall-MHD simulation described in the previous section ($d_i = 0.145$ m, $\eta_H = 10^{-8}$), and the other two are toroidal axisymmetric simulations with the same parameters, but with additional vertical fields of $B_V = -0.04 B_0 = -0.02$ T and $B_V = -0.06 B_0=-0.03$ T respectively. The initial condition for the toroidal simulation with vertical field of $B_V = -0.06$ is shown in the right panel of Figure~\ref{fig:init}.

 The initial centres of the current distributions are at $R=1.1$ m and $a=0.6$ m for all three simulations, but the centre of the O-points shifts slightly due to the addition of the vacuum fields. In the toroidal simulations, the flux-ropes oscillate radially due to the inbalance of the Lorentz forces from the vertical field, and the restoring hoop-force (there may also be forces due to image currents in the conducting walls). This radial motion of the flux-ropes could be minimised by a suitable choice of $B_V$. However, we do not do this here, as there is clear radial motion of the flux-ropes towards the central post in the experimental fast-camera images (see in~\citet{yamadaEPS}). These radial oscillations have only small effect on the rate at which the O-points move to the midplane in the toroidal simulations. In the Cartesian simulation, the O-point does not get to the midplane due to the formation of the central island that stalls the reconnection. Islands are also formed in the toroidal simulations, but they are quickly ejected as the symmetries present in the Cartesian simulatons are broken in toroidal geometry. The total merge time is $T_{merge}\approx 25 \tau_0$ for both toroidal simulations. 

\begin{figure}
\centering
\includegraphics[width=0.40\textwidth]{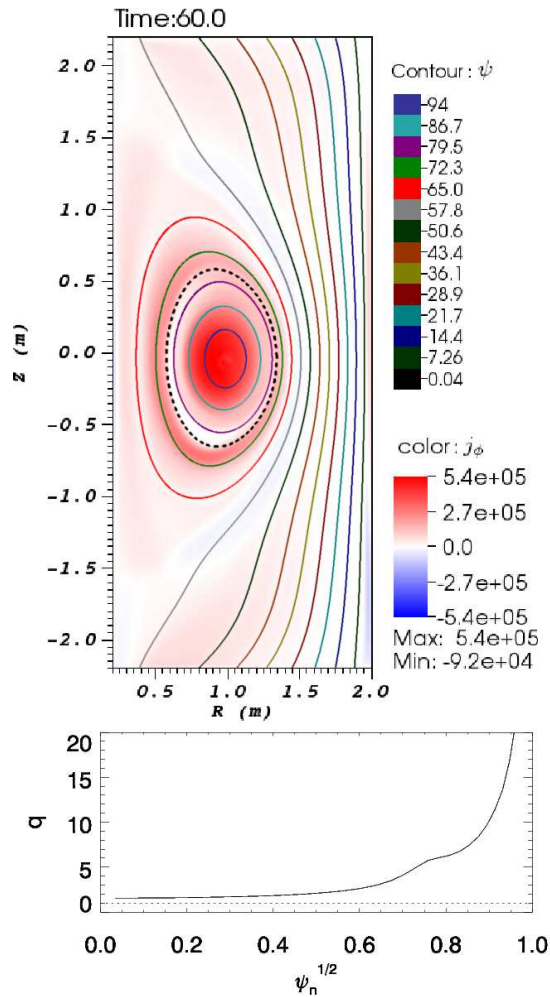}
\caption{\label{fig:tor-finalq}Top: Contours of the flux function, $\psi = RA_\phi$ in $\textrm{mWb}$, and toroidal current density, $j_\phi$ in $\textrm{A\,m}^{-2}$, in colour-scale after the two flux-ropes are fully merged. Below: The q-profile of the state at this time, plotted against the square-root of the normalised flux. }
\end{figure}

Figure~\ref{fig:tor-finalq} shows the flux, $\psi = R A_\phi$, and the toroidal current density, $j_\phi$, for the toroidal simulation with $B_v = -0.06 B_0$ at $t=60\tau_0$. At this time most of wave activity generated as the flux-ropes coalesce has died down and the final state relaxes to form a single spherical tokamak plasma of nested flux-surfaces. The dashed flux contour shown has the same value of flux as the one overlying the separator at $t=0$, see Figure~\ref{fig:init}). 

The bottom panel of Figure~\ref{fig:tor-finalq} gives the q-profile, or safety factor, for the magnetic configuration at $t=60 \, \tau_0$. This q-profile is a measure of the number of times the fieldlines loop around the vessel toroidally for each poloidal rotation, which is important for stability analysis of the tokamak plasma with respect to current driven instabilities. This q-profile is calculated by performing the integral 
\begin{equation*}q = \frac{1}{2\pi} \oint \frac{1}{R} \frac{B_\phi}{B_p} ds\end{equation*} 
around closed flux contours, where $B_p = \sqrt{B_R^2 + B_Z^2}$ and $ds$ is along the contour. It is plotted against the root of the normalised flux, defined by $\psi_n^{1/2} = \sqrt{\left[\psi(R_{\textrm{mag}}) - \psi(R)\right]/\left[\psi(R_{\textrm{mag}}) - \psi(R=0.2)\right]}$, where $R_{\textrm{mag}}$ is the position of the magnetic axis, and $R=0.2$ is the centre column which bounds the last closed flux surface.

A complete stability analysis of this final state is rather involved, and beyond the scope of this work. However, we do note that the safety factor is above the critical value of unity (the dashed line) for all values of $\psi_n^{1/2}$. For $q < 1$ the magnetic configuration may become unstable to the $m=n=1$ internal kink instability~\cite{wesson}, which would require 3D simulations to properly model.

\begin{figure}
\centering
\includegraphics[width=0.5\textwidth]{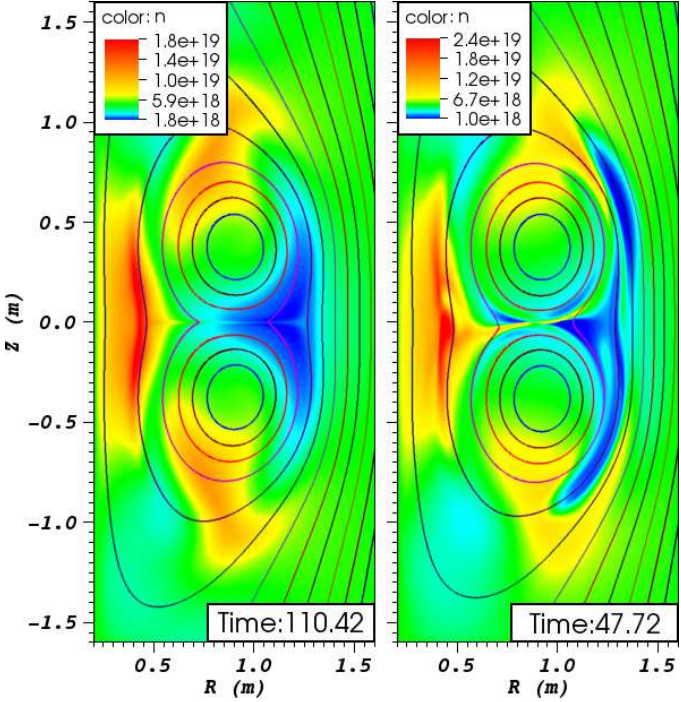}
\caption{\label{fig:tor-dens}Top: contours of the flux function $\psi = RA_\phi$, and number density, $n$ in $\textrm{m}^{-3}$, in colour-scale during the merging for the resistive $d_i=\eta_H=0$ (left), and Hall-MHD simulation $d_i=0.145$ m, $\eta_H=10^{-8}$ (right). }
\end{figure}

\begin{figure}
\centering
\includegraphics[width=0.5\textwidth]{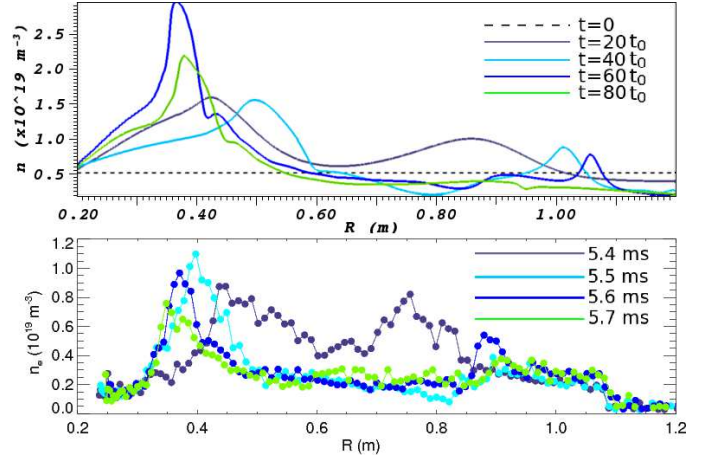}
\caption{\label{fig:traces}Top: Simulated Thomson scattering density traces from a toroidal axisymmetric Hall-MHD simulation at $Z=0.015$ m and $R\in[0.2,1.2]$ at $t=0, 20, 40, 60, 80 \tau_0$. Bottom: Electron density profiles from experiment measured by the Nd:YAG TS laser across the same radial chord at $t=5.5$ ms, $5.6$ ms, $5.7$ ms and $5.8$ ms.}
\end{figure}

Figure~\ref{fig:tor-dens} shows the plasma density for a resistive MHD simulation ($d_i=\eta_H=0$, top left) and a Hall-MHD simulation ($d_i=0.145$, $\eta_H=10^{-8}$, top right), both in toroidal axisymmetric geometry. The radius, $w$, and the peak current density, $j_m$, of the flux-ropes in the initial conditions are the same as in previous simulations, but now the initial position of the flux-ropes is at $R=0.9$ m, $Z=\pm a = \pm 0.8$ m. This was chosen so the flux-ropes merge closer to the central column, qualitatively matching the fast-camera images\cite{yamadaEPS}. The same amount of flux has been reconnected in both figures. In the resistive MHD simulation there is region of high density between the flux-ropes and the central post, and an $\mathcal{O}(1)$ density cavity on the outboard side. Following the density in time (not shown), we see the inboard (outboard) density increase (decrease) in both the ideal phase as the flux-ropes move towards the central post, as well as when there are strong reconnection outflows towards the inboard side (which has smaller volume due to the tight-aspect ratio toroidal geometry). Due to the low plasma-beta these density variations can be large, $\mathcal{O}(1)$, and equalise on timescales longer than the merge time. This effect has some similarity to a thermal pressure asymmetry, found in simulations of reconnection within toroidal geometry by~\citet{murphy08}.
The density plot for the Hall-MHD simulation appears more complex. However, it can be understood as the super-position of the toroidal resistive-MHD density asymmetry and the Cartesian quadrupole-like density asymmetry of Figure~\ref{fig:denscart}. A region of high density lies along the inner-lower outer-upper separator that results in the maximum density on the inboard side below the midplane, and gives a region of increased density within the outboard cavity above the midplane.

The 130 spatial point Nd:YAG Thomson scattering system installed on MAST\cite{scannell10} measures 1D profiles of electron temperature and density at extremely high resolution in both space and time. It is positioned to measure radial profiles at $Z=15$ mm above the geometric midplane (see Figure~\ref{fig:cartoon1}). Figure~13 of \citet{ono2012mast} shows profiles of electron temperature and density taken at a time resolution of $0.1$ ms during the merging. There is a double peaked profile in electron density at $5.4$ ms, that is typical of many merging-compressions shots. At later times the inner peak remains but the outer radial peak decays. The density profile for this shot has been reproduced here in Figure~\ref{fig:traces} (bottom panel) for a larger range of radial values.

The top panel in Figure~\ref{fig:traces} shows simulated radial density profiles at the same location, $R\in[0.2,1.2]$ m, $Z = 0.015$ m, as the Nd:YAG laser. These profiles are taken every $20 \tau_0$ during the merging for the Hall-MHD simulation. At $t=20 \tau_0$ there is a clear double peak in the density profile as the cut intersects the high-density region on the inboard side and the high-density separator arm on the outboard side. As the merging progresses, the radial position of the inner peak changes, depending on the position of the inner edge of the radially oscillating flux-ropes. The second peak is pushed radially outwards, as the flux-ropes collide, and over time it decreases in magnitude. At $t=80\tau_0$ (green line) the flux-ropes in the simulation are fully merged and the quadrupole-like density feature has disappeared, it is only present during reconnection. To explain this double peaked profile in the simulations, which shows similar evolution to the experimental profiles, needs both two-fluid effects and tight apect-ratio toroidal geometry. 

\section{\label{summary}Summary}

In this paper we have presented 2D fluid simulations of merging-compression plasma start-up within the Mega-Ampere Spherical Tokamak (MAST). In resistive MHD ($d_i = \eta_H = 0$), the flux-ropes enter the sloshing-regime due to magnetic pressure pile-up on the sheet edge for low resistivities ($\eta \lesssim 10^{-5}$). In the Hall-MHD simulations ($d_i = 0.145$ m, $\eta_H \neq 0$), the qualitative behaviour of the merging depends upon the ratio of the collisional current sheet width, $\delta$, to the ion-sound radius, $\rho_{is} = \sqrt{T_e/m_i}/\Omega_{ci} = \sqrt{\beta_e/2}\, d_i$. We varied $\delta$ by changing $\eta_H$, as the hyper-resistivity balances the reconnection electric field at the X-point. In the limit of $\delta \gg \rho_{is}$, the reconnection rate tends to the collisional limit. For $\delta \ll \rho_{is}$, the outflow separatrices open up and the peak reconnection rate increases, in agreement with previous studies~\cite{kleva95,simakov10,schmidt09}. However, in the intermediate regime, $\delta \gtrapprox \rho_{is}$, we find that the current sheet is highly unstable to a fast tearing-type instability, which is not present in the purely collisional case ($\rho_{is} = d_i = 0$) for the same dissipation coefficients. For $\eta_H = 10^{-8}$ a central island forms and stalls the reconnection.

In toroidal axisymmetric geometry, the flux-ropes oscillate radially between the hoop-force and additional vertical flux. This breaks symmetries present in the Cartesian case, and the central island can be ejected. The final state after merging and relaxation is a single Spherical Tokamak plasma with nested flux-surfaces. The density profiles in these toroidal Hall-MHD simulations are affected by the tight-aspect ratio toroidal geometry and two-fluid effects. Simulated line profiles of the plasma density against major radius, in a slice along the current sheet, show double-peaked profiles that have very similar time evolution to those seen in experimental Thomson Scattering profiles. 

Future work will examine the evolution of separate ion and electron temperatures in the toroidal axisymmetric simulations, to compare with the experimental Thomson scattering measurements of $T_e$ and possible future experiments that will measure 2D ion temperature distributions.   

\begin{acknowledgments}
This work was funded by STFC, the US DoE Experimental Plasma Research program, the RCUK Energy Programme under grant EP/I501045, and by the European Communities under the Contract of Association between EURATOM and CCFE. To obtain further information on the data and models underlying this paper please contact PublicationsManager@ccfe.ac.uk. The views and opinions expressed herein do not necessarily reflect those of the European Commission. 

Simulations were run at the Solar-Terrestrial Environment Laboratory (STEL) cluster at Nagoya University, and at the St-Andrews MHD cluster funded by STFC. A.S. would like to thank Kanya Kusano for access to the STEL computer, and also Grigory Vekstein, Takuma Yamada, Hiroshi Tanabe and Yasushi Ono for discussions related to this work.
\end{acknowledgments}

\appendix

\nocite{*}

\providecommand{\noopsort}[1]{}\providecommand{\singleletter}[1]{#1}%

\end{document}